\documentclass[useAMS,usenatbib]{mn2e}
\usepackage{graphicx,epsfig}
\usepackage{ulem}
\usepackage{amsfonts}

\def\be{\begin{equation}}
\def\ee{\end{equation}}
\def\ba{\begin{eqnarray}}
\def\ea{\end{eqnarray}}

\def\msun{M_\odot}
\def\lsun{L_\odot}

\def\ltsima{$\; \buildrel < \over \sim \;$}
\def\simlt{\lower.5ex\hbox{\ltsima}}
\def\gtsima{$\; \buildrel > \over \sim \;$}
\def\simgt{\lower.5ex\hbox{\gtsima}}
\def\etal{{et al.\ }}

\title[Far-infrared from growing BHs]
{Far-infrared nebular spectral features from growing massive black holes }
\author[E. O. Vasiliev \etal]
       {Evgenii O. Vasiliev$^{1,2}$\thanks{E-mail:eugstar@mail.ru},
        Yuri A. Shchekinov$^{2,3}$, Biman B. Nath$^3$ \\
$^1$Southern Federal University, Stachki Ave. 194, Rostov-on-Don, 344090 Russia\\
$^2$Lebedev Physical Institute of Russian Academy of Sciences, 53 Leninskiy Ave., 119991, Moscow\\
$^3$Raman Research Institute, Sadashivanagar, Bengaluru 560080, Karnataka, India
}
\begin{document}
\date{Accepted 2018 December 32.
      Received 2018 December 31;
      in original form 2018 December 31}
\pagerange{\pageref{firstpage}--\pageref{lastpage}}
\pubyear{2119}
\maketitle

\label{firstpage}

\begin{abstract}
Supermassive black holes (BHs) and their host galaxies are interlinked by virtue of feedbacks and are thought to be co-eval across the Hubble time. This relation is highlighted by an approximate proportionality between the BH mass $M_\bullet$ and the mass of a stellar bulge $M_\ast$ of the host galaxy. However, a large spread of the ratio $M_\bullet/M_\ast$ and a considerable excess of BH mass at redshifts $z\sim8$, indicate that the coevolution of central massive BHs and stellar populations in host galaxies may have experienced variations in its intensity. These issues require a robust determination of the relevant masses (BH, stars and gas), which is difficult in the case of distant high-redshift galaxies that are unresolved. In this paper, we seek to identify spectral diagnostics that may tell us about the relative masses of the BH, the gas mass and stellar mass. We consider general features of SEDs of galaxies that harbour growing massive BHs, forming stars and interstellar/circumgalactic gas. We focus on observational manifestations of possible predominances or intermittent variations in evolutionary episodes of growing massive BHs and forming stellar populations. We consider simplified scenarios for star formation and massive BHs growth, and simple models for chemical composition of gas, for dust free gas as well as for gas with dust mass fraction of $1/3$ of the metal content. We argue that wideband multi-frequency observations (X-ray to submillimeter) of the composite emission spectra of growing BH, stellar population and nebular emission of interstellar gas are sufficient to infer their masses.
\end{abstract}

\begin{keywords} 
Accretion, accretion discs --- black hole physics --- {\it galaxies:} quasars: supermassive black holes  --- ISM: general --- infrared: ISM --- submillimeter: ISM   
\end{keywords}


\section{Introduction}

\noindent

All massive galaxies in local universe are 
believed to host supermassive black holes (SMBHs) in their centres. Moreover, the mass of SMBHs $M_\bullet$ are found to  correlate with the mass of stellar population $M_\ast$ of host galaxies \citep[][]{marconi03,haring04,sani11,kormen13,heckman14}, which suggests a synergy between the formation and growth of SMBHs and their host galaxies. The $M_\bullet\hbox{--}M_\ast$ correlation  is however ambiguous. The dominant source of uncertainty stems from measurements of the SMBH mass, because most observational techniques require a calibration of $M_\bullet$ to the observables used to measure it \citep{bentz13,kormen13,woo13,heckman14}. Further, the determination of stellar mass relies on the `mass-to-light' ratio, which is only known within an accuracy of $\sim 10$\% \citep{kauf03}. For late-type low-mass galaxies ($M_\ast<10^{11}\msun$), there can also be uncertainties arising from strong starbursts during the last 1 Gyr \citep{salim05}. As a result, {the relation between} SMBH mass $M_\bullet$ {and} stellar mass $M_\ast$ {exhbits} a spread of more than an order of magnitude \citep[see Fig 9b in][]{heckman14}. 
{This has led}  \cite{heckman14} {to conclude} that the correlation between $M_\bullet$ and $M_\ast$ in the local universe cannot be approximated as a simple linear proportionality. 
{These complications}  reflect the nature of mutual feedbacks at work in the process of star formation and a growing SMBH in a host galaxy. At the same time, the analyses by \cite{graham13,woo13} and \cite{kormen13} indicate that $M_\bullet$ and stellar velocity dispersion in galactic centres $\sigma$ are coupled tighter than $M_\bullet\hbox{--}M_\ast$. This  implies that black hole growth is regulated mainly by the galaxy gravity, rather than by feedbacks between the growing SMBH and the stellar population, and it is essentially important  to identify the contributions of all variables in this interplay. However, given the observational uncertainties, it is challenging to disentangle the mutual influences of growing BHs, the assemblage of the host galaxy itself, and the formation of stellar population in it during these epochs. 

The problem regarding the $M_\bullet-M_\ast$ relation becomes particularly acute when host galaxies at $z\simeq 7$ are concerned. Observations of [CII] 158 $\mu$m line in a set of $z\simgt 6$ quasars have led \cite{volont16} and \cite{decarli18} to conclude that the ratio  $M_\bullet/M_\ast$ at $z\simeq 6-7$ epochs is more than an order of magnitude higher than that in the local Universe. While at redshifts $z> 6$, the mean ratio of the BH to the {\it dynamical} mass is $\langle M_\bullet/M_{dyn}\rangle\simgt 0.02$ for sufficiently strong (${\rm S/N}\geq 10$) [CII] lines \citep[see Fig. 12 in~][]{decarli18}, at low redshifts $z\sim 0$ the BH to {\it stellar} mean mass ratio is only   $\langle M_\bullet/M_\ast\rangle\simgt 0.003$ when host galaxies with AGNs are concerned \citep[see Fig 10 left in~][]{reines15}. 
Under normal circumstances, the ratio of BH mass to dark matter mass would have been smaller in comparison with its ratio to stellar mass. 
The observed ratios, which are the other way around, perhaps reflect the dynamical features of mutual influences of growing BHs and stellar populations of host galaxies on the initial episodes or during the entire evolution. Examples of such models with $M_\bullet/M_\ast$ up to $\sim 0.1\hbox{--}1$ are described in \cite{agarwal13,volonterid16}. However, observational selection effects cannot be  neglected in such observations \citep{decarli18}, and complementary direct observations would be crucially important for unveiling the synergy between SMBHs and their hosts.  

{It therefore appears that there may have been periods during the co-eval evolution of SMBH and its host galaxy in which the black hole mass may dominate over stellar and gas mass. It then becomes important to ask if this idea can be tested through observations. Although the masses of BH and stars can be estimated from the spectra of galaxies, such estimates become difficult in the presence of gas and dust.}
{Previous works have identified some} spectral features in optical, near- and mid-infrared bands {for the case of} SMBHs surrounded by interstellar medium of low-metallicity ($Z\leq 0.05~Z_\odot$) {based on a few scenarios} \citep{agarwal13,nataraj17,volont17}. One scenario implies a $\sim 100~\msun$ black hole remnant of a massive Population III star as a SMBH seed \citep[][]{tanaka09,madau14,volonteri15,lupi16,pezzulli16}, while in the other a more massive direct collapse black hole (DCBH) of $M\sim 10^5~\msun$ acts as a seed. In both cases the stellar population is supposed to remain underdeveloped, such  that growing BHs dominate the production of photons\footnote{\cite{agarwal13} named such stages of galactic evolution as OBG -- Obese Black hole Galaxies.}. Spectral lines in mid-infrared (MIR) range turn out to be sensitive to heating and ionization regimes, which makes them possible in principle to {identify}  with the upcoming {\it James Webb Space Telescope} ({\it JWST}). However, scenarios with Pop III black hole seeds predict the MIR line intensities to be weaker than the {\it JWST} detection limit \citep[][]{valiante16,valiante18}, which makes it difficult to discriminate between the two scenarios, requiring alternative approaches  to be pursued. 

\begin{figure}
\center
\includegraphics[width=8cm]{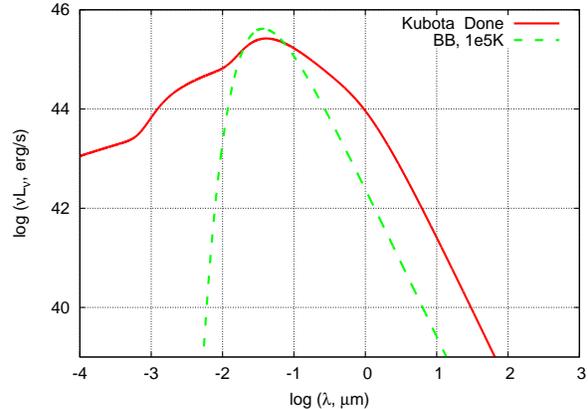}
\caption{
Spectral energy distributions for the \citet{kubota18} model for {$M_\bullet=10^8\msun$}
 and {a black body with} $T=10^5$~K.
 }
\label{fig-spec-bh}
\end{figure}

In the current paper we {address the question whether observations of spectral energy distributions (SED) consisting of contributions from SMBHs,  stellar populations and galactic gas of a distant host galaxy can be disentangled to infer the masses of these three actors. For this purpose we explore the multi-wavelength spectra: X-ray, optical, infrared and far-infrared wavebands} to probe the synergy between growing massive BHs and their host galaxies {when gas and dust shroud the central BH and stellar population}. More specifically, we address the question whether the {interrelations between the masses of black hole $M_\bullet$, of galactic gas $M_g$, and galactic stellar population $M_\star$ can be inferred through observations of the multi-wavelength spectra from such sources.  In order not to increase the number of free parameters,  we do not consider an {\it evolutionary} scenario. We focus only on whether the observed emissions from a galaxy consisting of stars, gas and a supermassive black hole can be discriminated in order to measure the masses of the constituents if the galaxy is not spatially resolved.}  

The paper is organized as follows. Section 2 contains the model description and numerical setup,  Section 3 describes the results, while in Section 4 we discuss observational diagnostics, Section 5 summarizes the results.


\section{Model description}
\subsection{Spectral energy distribution of a SMBH}
Following \cite{nataraj17} and \cite{kubota18} we assume that SMBH radiation comes from a slim disc \citep{abram88}, with the Novikov-Thorne spectrum \citep{novik73} modified by a Comptonised extention up to high ($E>1$ keV) energies \citep[see for more detail in][]{kubota18}. This choice is motivated by the fact that slim discs are radiatively inefficient and that the accretion flow is stable against Rayleigh-Taylor instability  even for super-Eddington accretion rate \citep[see discussion in][]{abram88,sadow09,pacucci15}. An example of the spectrum is shown in Figure~\ref{fig-spec-bh}. For comparison, we also show a blackbody spectrum, which illustrates that in longer wavelengths (infrared to submm) the BH spectrum can be approximated as thermal radiation with effective temperature depending on the BH mass $T_{eff}\propto M_\bullet^{1.5}$ \citep[see~][]{kubota18}.

\begin{figure}
\center
\includegraphics[width=8cm]{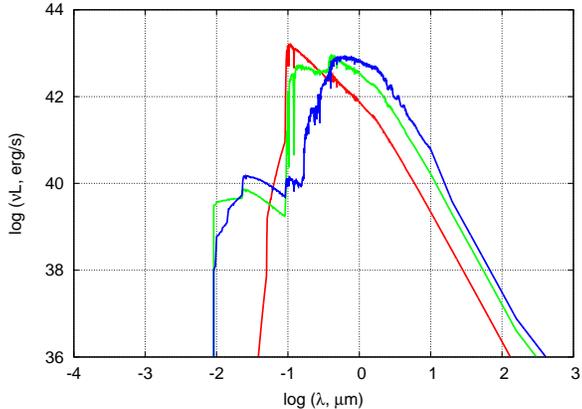}
\caption{
Spectral energy distributions for stellar population with ages of 30~Myr for ${\rm [Z/H]} = -2$ {(red line),  of 200~Myr for ${\rm [Z/H]} = -1$ (green line) and 1~Gyr for solar metallicity (blue line)} \citep{bruz03}. The bolometric luminosity is $L=3.1\times 10^9\lsun$. 
}
\label{fig-spec-glx}
\end{figure}

\subsection{Stellar population}
For the host galaxy we assume a spectral energy distribution (SED) of a single stellar population (SSP) based on the Padova 2000 stellar evolutionary tracks \citep{padova2000} with the Salpeter inital mass function (IMF) \citep{bruz03}. For SSP we consider two models with a low (${\rm [Z/H]} = -2$, $-1$) and with solar metallicity. The bolometric luminosity of the SSP is taken to be $L=3.1\times 10^9\lsun$ and $3.1\times 10^{10}\lsun$, equal to Eddington luminosities of BHs with $M_\bullet = 10^5~\msun$ and $10^6~\msun$, respectively. The corresponding stellar masses can be roughly estimated as $M_\ast\sim 10^{10}$ and $M_\ast\sim 10^{11}\msun$ for local late-type galaxies \citep[see in~][]{faber79}. 

{Figure~\ref{fig-spec-glx} presents three SEDs models of stellar population from the spectra library of \citet{bruz03}: a young population of an age 30~Myr for  ${\rm [Z/H]} = -2$, a middle-aged (200~Myr) population for ${\rm [Z/H]} = -1$ and a moderately old population of 1~Gyr for solar metallicity.}  Increasing the  metallicity for the young SSP leads to an enhanced luminosity at the IR/submm wavelengths. Older populations show two noticeable features in the SEDs. On one hand, in the initial 100~Myr young massive stars dominate in the UV part of the spectrum, and at later epochs, the most massive stars leave the main sequence and evolve into red giants, making the UV part to decline and the infrared emission to rise. On the other hand, low-mass stars at later stages $\simgt 120\hbox{--}150$~Myr evolve through the post-AGB phase and maintain far-UV emission up to several Gyr. 

We are interested in the full spectrum consisting of  the cumulative intrinsic spectra of the stars and the SMBH, transmitted through the interstellar gas, and the nebular emission of the ISM ionized and heated by these sources. Our aim is to determine how the spectral features--- in the continuum and in lines--- in the IR/submm range are connected with the BH and stellar masses, as well as the ISM gas mass (with and without dust). Our hope is that such a connection will allow us to infer a possible correlation between them.

\subsection{CLOUDY Setup}
Ionization states of various elements and the ionic spectral lines are calculated with the CLOUDY code \citep{cloudy17}. The gaseous layer is exposed to radiation from the central sources, i.e. a BH and stellar population. We assume spherical geometry, in which the inner radius of the layer is fixed at $100$~pc, and the outer radius is determined by the gas mass $M_g$. The density and metallicity are distributed homogeneously; calculations are done for {three gas metallicities: ${\rm [Z/H]}= -2,\  -1$ and 0,  corresponding to the stellar population models presented in Figure~\ref{fig-spec-glx}}, the fiducial density is $n=1$ cm$^{-3}$. A decrease in the  metallicity leads to a nearly proportional decrease of luminosities in metal lines. In runs with gas density different from the fiducial value, the gas mass is kept fixed. We assume that the mass of the layer is fixed, such that the outer radius of the layer varies as $n^{-1/3}$. We also assume that the gas is in  thermal equilibrium.

\begin{figure*}
\center
\includegraphics[width=16cm]{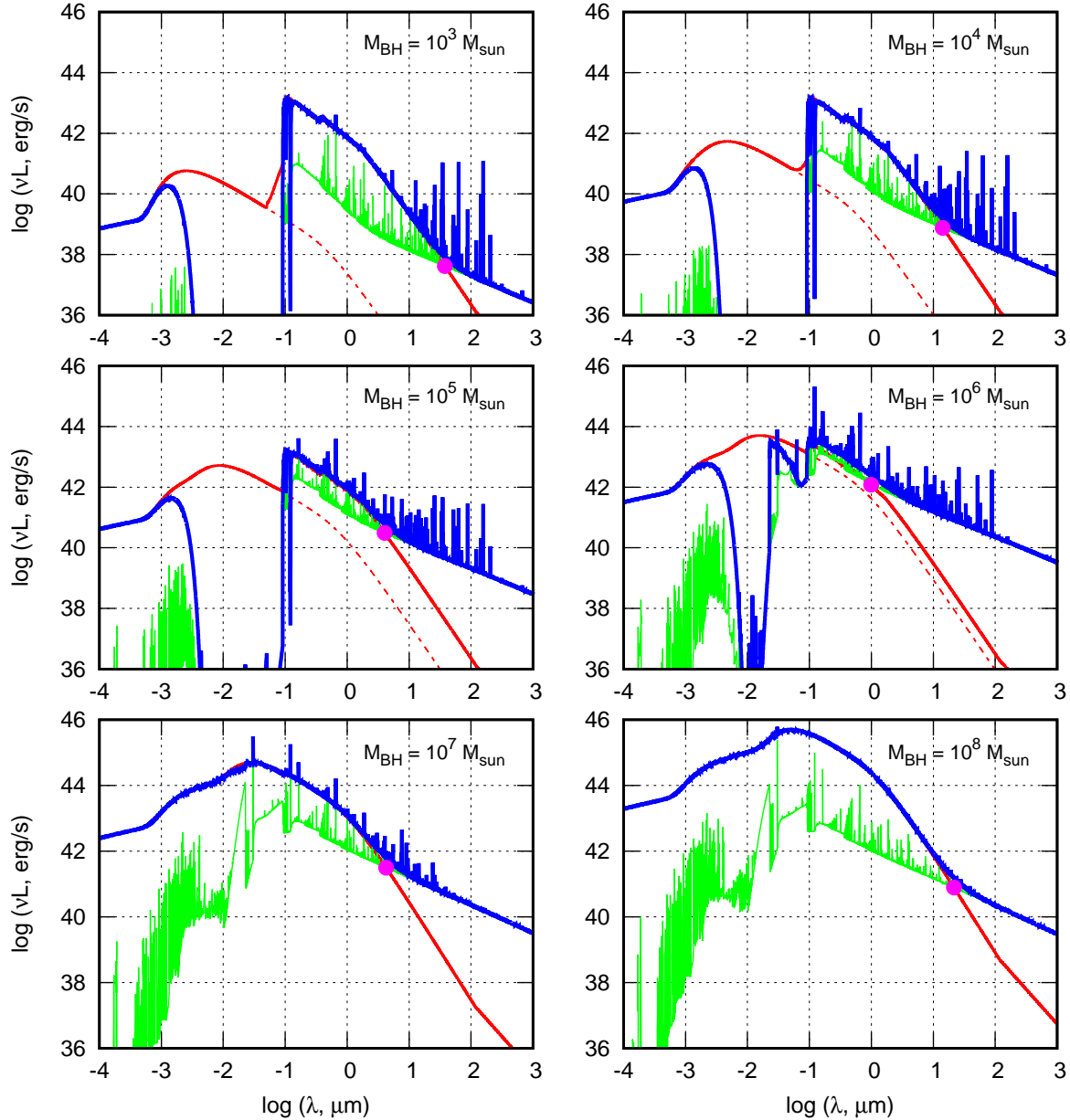}
\caption{
Emission spectrum of diffuse interstellar gas photo-ionized by radiation field from a single stellar population and  growing BH with mass $M_\bullet = 10^3, 10^4, 10^5, 10^6, 10^7$ and $10^8~\msun$, respectively. The SSP bolometric luminosity is equal to $L=3.1\times 10^9\lsun$ and the age is 30~Myr, the metallicity is ${\rm [Z/H]} = -2$. The SSP spectum is presented in Figure~\ref{fig-spec-glx}. A slim-disc spectrum is assumed for the BH \citep{kubota18} with the Eddington bolometric luminosity; gas mass is $M_g=10^{10}~\msun$. Thin {\it red  lines}  depict the sum of stellar plus BH (shown separately  by {\it red dashed lines}) incident flux, {\it blue lines} show the net transmitted spectrum -- the sum of an attenuated incident flux plus diffuse continuum and line emission, {\it green lines} show the emission spectrum from photo-ionized and heated diffuse interstellar gas. Interstellar density is $n=1$~cm$^{-3}$, gas metallicity $[{\rm Z/H}]=-2$. 
}
\label{fig-spec-sspbh}
\end{figure*}

\begin{figure}
\center
\includegraphics[width=8cm]{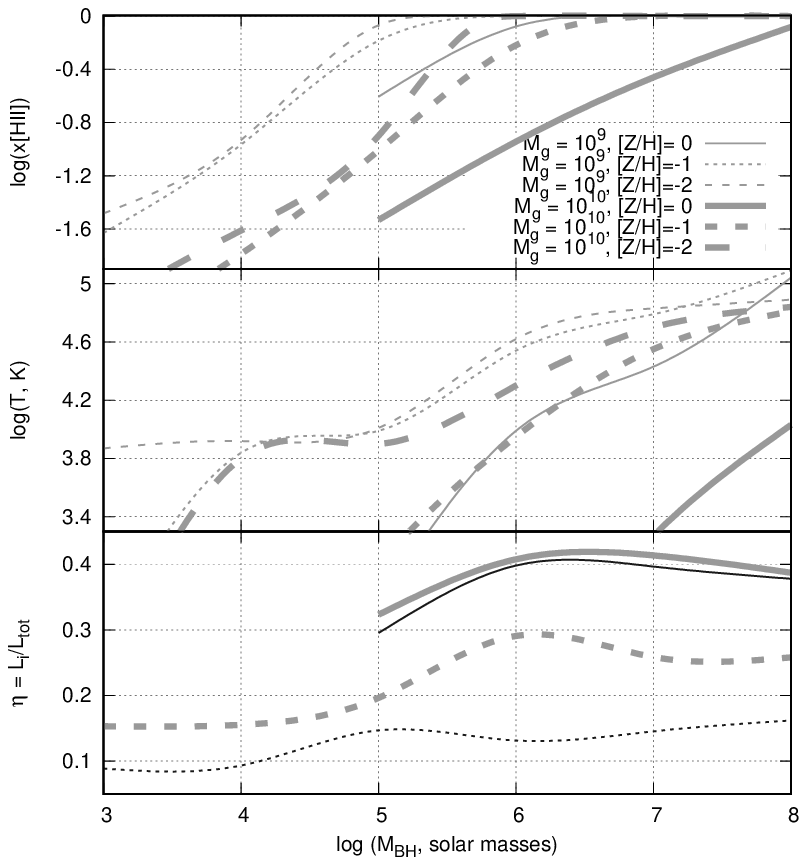}
\caption{
HII fraction ({\it upper panel}), electron temperature ({\it middle panel}) in gas layer with mass $M_g = 10^9~\msun$ (thin lines), $10^{10}~\msun$ (thick lines) {and metallicity ${\rm [Z/H]}=-2$ (dashed lines), $-1$ (dotted lines), 0 (solid lines)} exposed to the radiation field from a SSP with bolometric luminosity equal to $L_*=3.1\times 10^9\lsun$ and a growing BH with mass $M_\bullet$ indicated on $x$ axis; {for ${\rm [Z/H]}=-1$ and 0} dust is assumed to present with mass fraction ${\cal D}=0.3\zeta$. {{\it Lower panel} depicts the fraction of energy spent by stars and BH in heating dust (Sec. \ref{dst_ht}), as shown by thin and thick solid lines, for the gas mass $M_g=10^{9}$ and $10^{10}\msun$ respectively.}  
}
\label{fig-xe-te}
\end{figure}


\section{Results}
{In low-metallicity cases ($\rm [Z/H]\leq-2$), we consider for illustrative purposes a wide range of BH masses $10^3\leq M_\bullet\leq 10^8\msun$. Even though their contribution in heating and ionizing galastic gas in the low BH mass end $10^3\leq M_\bullet\leq 10^5\msun$ is not significant (see below  in Figure \ref{fig-spec-sspbh}), they may reveal anticipated interrelations between emission characteristics of galaxies with BHs at their initial growth episodes. However, when galaxies with higher metallicity  are concerned, the effects of interstellar opacity come into play and complicate the calculations. Therefore, in models with high metallicity we limit our calculation to BH masses $10^5\leq M_\bullet\leq 10^8\msun$, because BHs with masses $M_\bullet\geq 10^5\msun$ are able to ionize absorbing species (carbon, oxygen and nitrogen) in ambient interstellar gas and make it more transparent. { Higher gas transmittance also prevents the BHs from producing dynamical (gravitational) feedback effects on the very inner gas and dust layers.}} 

\subsection{Dust-free gas}
Let us first consider the combined effect of the stellar and BH radiation on emissivity of galactic gas with a sub-solar metallicity $\rm [Z/H]=-2$.  We consider a set of models with a fixed mass of stellar population corresponding to bolometric luminosity $L_b=3.1\times 10^9L_\odot$, complemented by radiation from an accreting BH with masses from $10^3\msun$ to $10^8\msun$; the mass of a surrounding interstellar gas is assumed $M_g=10^{10}\msun$. The total radiation from stars and the BH irradiates the diffuse gas, resulting in its ionization, heating and consequently affecting  its emissivity. 

Figure~\ref{fig-spec-sspbh} presents for a low metallicity gas the sum of the stellar and the BH incident flux ({\it red lines}), the continuum and line emission spectrum from photo-ionized  diffuse interstellar gas ({\it green lines}), and the net transmitted spectrum ({\it blue lines}). It is seen that, regardless of the low metallicity ${\rm [Z/H]} = -2$,  UV and soft X-ray photons produced by BH are absorbed in a partially neutral gaseous layer, with a fraction of the absorbed energy being converted to IR/sub-mm lines of hydrogen and low ionization states of metals, such as [NII]~121~$\mu$m, [CII] 158~$\mu$m, [NII] 205~$\mu$m and others. However, the most important effect of these photons is an increase in heating and ionization in the gas layer. This manifests in an enhanced free-free nebular continuum. 

It is readily seen that gas nebular emission is predominantly due to hard radiation from the BH. As the BH mass increases the nebular luminosity increases proportionally. When the ionization fraction approaches unity (Figure~\ref{fig-xe-te}), this proportionality in continuum emission weakens, while line emission decreases. The corresponding decrease of the neutral fraction makes the outer gas layer more transparent, as can be seen from  comparing the mid-left and mid-right panels on Figure~\ref{fig-spec-sspbh} for BH mass $M_\bullet \simgt 10^5$ to $M_\bullet \simgt 10^6~\msun$. For higher BH masses (two bottom panels), Hydrogen is completely ionized, and the incident SED remains practically unchanged after passing through the gas layer. Further increase of BH mass above $M_\bullet>10^6\msun$ results in decreasing the intensities of H, He and metal lines. Note that for the same BH mass limit, $M_\bullet \simgt 10^6~\msun$, the BH outshines the stellar radiation from the SSP with bolometric luminosity $L_*=3.1\times 10^9\lsun$. 

The luminosity of free-free nebular emission in the IR/submm range increases as $L_\nu\propto M_\bullet$ until the BH masses exceeds a critical value $M_{\bullet,s}$, and afterwards it saturates and remains nearly constant at higher $M_\bullet$. Assuming that gas in the layer is ionized mostly by EUV and X-ray photons from the BH one can estimate its mass when free-free emission saturates as the BH mass crosses the limit, as follows,
\be \label{mbcrit} 
M_{\bullet}>M_{\bullet,s}\sim {\alpha_rn\langle \epsilon_i\rangle M_g\over m_pL_{Edd, M_\odot}}\simeq 10^{-4}nM_g \,,
\ee 
where $\alpha_r\simeq 2.3\times 10^{-13}T_4^{-0.82}$ cm$^3$~s$^{-1}$, the case B hydrogen recombination rate \citep{draineb}, $\langle \epsilon_i\rangle=\int F_\epsilon\sigma_i(\epsilon)\epsilon d\epsilon/\int F_\epsilon\sigma_i(\epsilon) d\epsilon\sim 0.05$ keV is the mean energy of ionized photons, $F_\epsilon$ is the energy flux of X-ray photons, {$L_{Edd, M_\odot}=1.26\times 10^{38}$} erg~s$^{-1}$ is the Eddington luminosity for a $M_\bullet=\msun$ black hole, $n$ is gas density in the layer, $M_g$ is its mass. We assume $\sigma_i\simeq 2\times 10^{-22}\epsilon_{\rm keV}^{-2.5}$ cm$^2$ for H + He plasma \citep{wilms00}. 

For a gas mass of $M_g=10^{10}\msun$ Eq. (\ref{mbcrit}) gives $M_\bullet\sim 10^6\msun$. The fractional ionization  $x\simlt 1$, as seen in Figure \ref{fig-xe-te} and further growth of $M_\bullet$ does not increase the luminosity of nebular emission. Assuming for H + He plasma the cooling rate $\Lambda_{\rm neb}\simeq 3\times 10^{-25}T_4^{1/2}$ erg~cm$^3$~s$^{-1}$ -- the sum of free-free and free-bound emission \citep{draineb}-- one can estimate the fraction of bolometric BH luminosity being radiated as nebular emission,    
\be 
\eta_{\rm neb}\sim 3\times 10^{-6}x^2nT_4^{1/2}{M_{g}\over M_\bullet} \,.
\ee 
For $M_\bullet=10^6$, this fraction is $\eta_{\rm neb}\sim 0.3$ for fiducial parameters, and it decreases with an increase in $M_\bullet$. 

\begin{figure}
\center
\includegraphics[width=8cm]{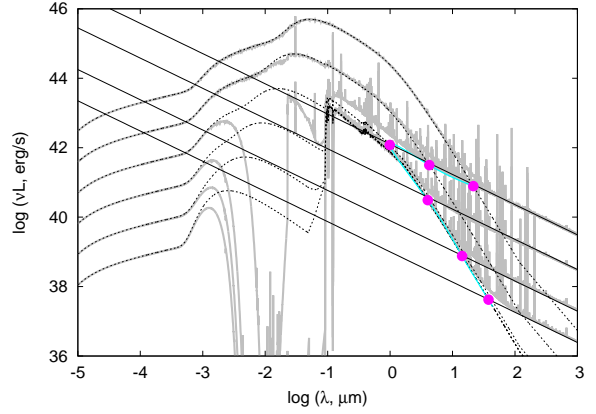}
\caption{
A schematic presentation of the spectra shown in Figure~\ref{fig-spec-sspbh} (all transmitted spectra in one plate, the emission spectra from photo-ionized diffuse interstellar gas are replaced by lines with slopes followed by the free-free continuum). The thick dark grey line depicts the SSP incident spectrum. The set of thick light grey lines present the total transmitted spectra for the SSP+BH source. The straight thin lines follow pure free-free continuum. Color symbols depict the location of intersection points. 
}
\label{fig-spec-scheme}
\end{figure}

In the long-wavelength limit, the free-free nebular spectral luminosity $L_\nu^{\rm n}\propto \nu^{-0.12} \propto \lambda^{0.12}$. This scaling is shallower than that of the sum of stellar and BH spectra,  $L_\nu^{\rm\small SSP+BH}\propto\nu^2\propto\lambda^{-2}$. Therefore, at a certain wavelength $\lambda_i$, the nebular emission outshines the radiation from the original sources. At $\lambda_i$, the net transmitted spectrum changes from the slope $\alpha=-2$  shortward to $\alpha=-0.12$ longward. In Figure~\ref{fig-spec-sspbh}, we show  the spectra in the form of ${\cal L}_\nu\equiv\nu L_\nu$, which changes from ${\cal L}_\nu\propto\nu^3 \propto\lambda^{-3}$ in the short-wavelength side $\lambda<\lambda_i$ to ${\cal L}_\nu\propto\nu^{0.88} \propto\lambda^{-0.88}$ after the inflection point, located in the far-IR or the sub-mm wavelength range depending on masses of SPP, BH and the ambient gas layer. The presence of such an inflection point in spectrum of a distant source would indicate the prevalence of the nebular emission. 

Magenta points in Figure~\ref{fig-spec-sspbh} depict the intersection between the free-free continuum and stellar/BH thermal background. One can observe that for a given SSP luminosity an increase of BH mass leads to a shift of the intersection towards shorter wavelengths with simultaneous increase of the luminosity, until the SSP dominates in the optical and IR range -- it is seen in the four upper  panels of Figure~\ref{fig-spec-sspbh}, for BH masses from $10^3$ to $10^6~\msun$. For more massive and brighter BHs the free-free continuum saturates and further increase of the BH mass shifts the inflection point to longer wavelengths with simultaneous decrease of luminosity. 

\begin{figure}
\center
\includegraphics[width=8cm]{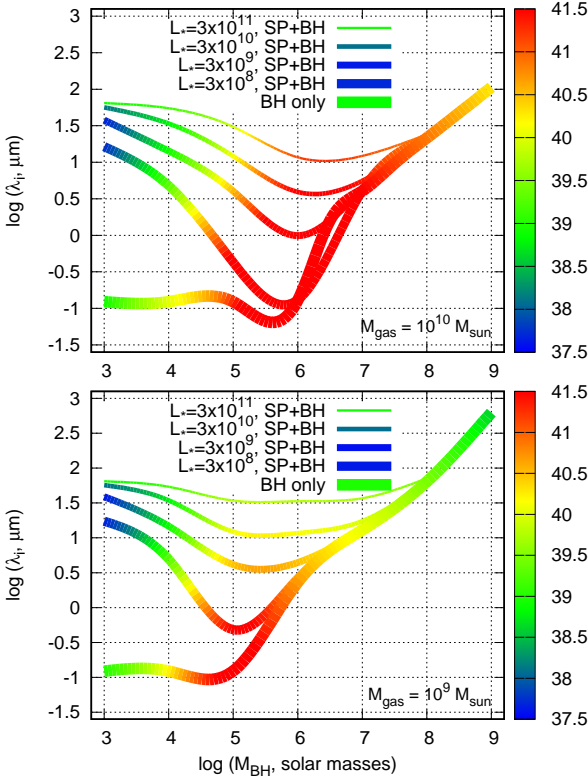}
\caption{
The wavelength of the intersection (magenta points in Figure~\ref{fig-spec-sspbh}) between the free-free continuum by the photoionized gas and the net transmitted spectrum -- the sum of an attenuated incident flux plus diffuse continuum and line emission versus the BH mass for four values of stellar population luminosity: no stellar population (the thickest line, 'BH only'), $L_*=3.1\times 10^8,\ 3.1\times 10^9,\ 3.1\times 10^{10},\  3.1\times 10^{11}\lsun$ (from thicker to thinner lines). The SSP age is 30~Myr. The gaseous mass of the layer exposed to the radiation is $M_g = 10^{10}\msun$ (upper panel), $10^{9}\msun$ (lower panel). The color along the lines correspond the logarithm of luminosity $\nu L_\nu$ reached at the intersection point.
}
\label{fig-spec-int}
\end{figure}

For clarity, we collect all spectra in a schematic plot on Figure~\ref{fig-spec-scheme}. The thick dark grey line depicts the SSP incident spectrum. A set of thick light grey lines presents the total transmitted spectra of SSP+BH source with increasing values of $M_\bullet$. The straight thin lines follow pure free-free continuum. Colour symbols depict the location of intersection points. One can note that there is a turning point for the BH mass value $M_\bullet \sim 10^6~\msun$, when the free-free emission saturates. Another representation of such loop-like behaviour of the inflection wavelength on the mass of an illuminating BH is given in Figure~\ref{fig-spec-int}. 
 
The colours on lines in this figure correspond to the luminosity ${\cal L}_\nu$ at the inflection point as coded in the colour bar. The BH mass is a two-valued function of $\lambda_i$ and the luminosity ${\cal L}_\nu$. The dropping branch of $\lambda_i(M_\bullet)$ reflects that SPP optical and IR emission outshines the BH emission. On the right hand side of the Figure, $\lambda_i(M_\bullet)$ turns to grow when this relation reverses, approximately at $M_\bullet>M_{\bullet,s}$. 

\subsubsection{Spectral lines}\label{xsubmm}
Nebular spectral emission is fed  by both UV from stars and by X-ray and UV from the growing black hole. As mentioned above, the contribution from young stellar population (SSP) is restricted by a very narrow UV band $\lambda=500\hbox{--}1000$\AA\, from OB stars. As a result, in optical and near infrared range, the emission from stars dominates and nebular line emission remains weak. This is clearly seen in Figure \ref{fig-spec-sspbh}: when the BH mass lies within $M_\bullet< 10^5\msun$ all optical and even a fraction far IR nebular lines fall below the stellar emission. Only from $M_\bullet\geq 10^5\msun$ onward, the nebular lines emerge over and above the stellar continuum due to a predominant contribution of the BH towards ionization. However, further increase of the BH mass $M_\bullet > 10^6\msun$ results in a higher ionization and in a suppressed hydrogen recombination optical emission. Ultimately at $M_\bullet\geq 10^7\msun$ nebular emission becomes saturated. 

\begin{figure*}
\center
\includegraphics[width=17cm]{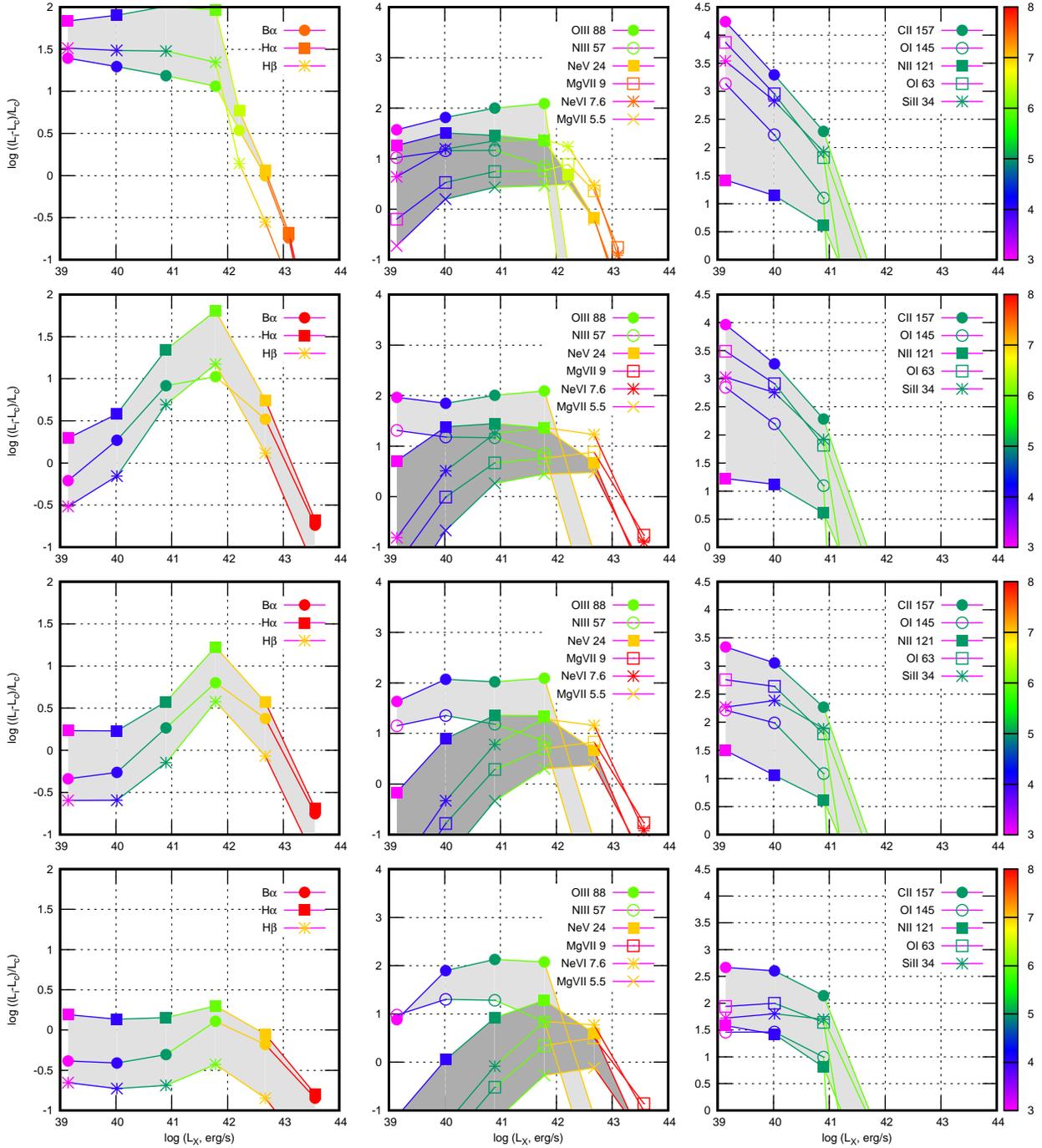}  
\caption{
Relative intensities of different emission lines shown versus the black hole X-ray luminosity in the 0.1--3~keV band: {\it left} column depicts hydrogen recombination lines (H$\alpha$, H$\beta$ and B$\alpha$), {\it middle} column shows Mid IR and IR lines of highly charged ions, and {\it right} column shows far IR lines of neutrals and singly charged ions. Panels from top downward depict relative intensities for gas illuminated only by a central black hole ({\it upper} row), SSP with $L_\ast=3\times 10^9L_\odot$, with $L_\ast=3\times 10^{10}L_\odot$, and with $L_\ast=3\times 10^{11}L_\odot$, correspondingly; gas mass in all panels is $M_g=10^{10}\msun$. The color bar marks logarithmic values of black hole masses corresponding to the Eddigton luminosity. }
\label{fig-xlines-rsmet}
\end{figure*}

We illustrate this  in Figure \ref{fig-xlines-rsmet} where relative intensities of nebular emission lines ${\cal F}_\lambda=(F_l(\lambda)-F_c)/F_c$, are shown against the black hole X-ray luminosity in the 0.1--3 keV energy band. Here $F_l(\lambda)$ is the line-centre intensity, $F_c$ is the net continuum intensity on both sides of the line. The upper row shows the dependences of relative line intensities when the gas layer is illuminated only by a BH with different masses. From left to right, optical hydrogen recombination lines H$\alpha$, H$\beta$ and B$\alpha$, mid- and far-IR {high ionization ions lines:} [MgVII] 5.5, 9 $\mu$m, [NeVI] 7.6 $\mu$m, and [NeV] 24 $\mu$m, [NIII] 57 $\mu$m, [OIII] 88 $\mu$m respectively, in middle {column} panels, and far-IR lines of neutrals and low-ionization species: [SiII] 34 $\mu$m, [OI] 63 $\mu$m, [NII] 121 $\mu$m, [OI] 145 $\mu$m, [CII] 158 $\mu$m in the right column panels, are presented. All lines with similar behaviour are joined by grey shade in order to clearly illustrate their similarity. It is readily seen that when the gas is illuminated only by a BH, the Hydrogen lines do not  show an increase in intensity with increasing BH mass until $M_\bullet\simlt 3\times 10^6\msun$ {(corresponding to $L_x \simlt 10^{42}$ erg s$^{-1}$}). Nearly the same dependence is shown by [NeV], [NIII], [OIII] in middle {column} panel. Intensities of ions with mid-IR lines [MgVII] and [NeVI] reveal a gradual growth with the BH mass (${\cal F}_\lambda\propto M_\bullet^{1/3}$) for $M_\bullet\simlt 3\times 10^6\msun$. In contrast, far-IR lines of low ionization and neutral species: [SiII], [OI], [NII], [OI], [CII] (right panel) drop quickly, as ${\cal F}_\lambda\propto M_\bullet^{4/3}$, and practically disappear at $M_\bullet\simgt 3\times 10^6\msun$.  

\begin{figure*}
\center
\includegraphics[width=16cm]{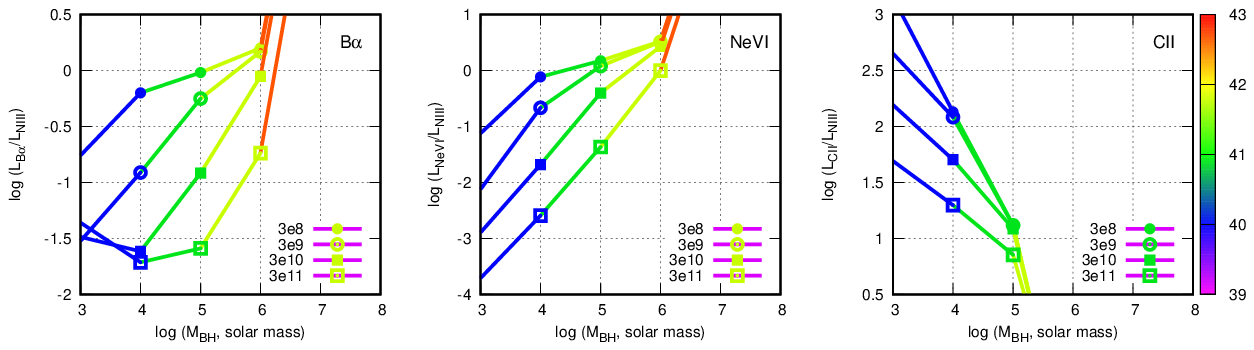}
\caption{
The line ratios -- relative intensities of the lines B$\alpha$, [NeVI] and [CIV] to the intensity of the line [NIII] -- left to right shown versus the BH mass, for different luminosities of the stellar population from $L_\ast=3\times 10^8L_\odot$ to $3\times 10^{11}L_\odot$ as legended. Lines are color-coded according to the BH X-ray luminosity according to the color bars. }
\label{fig-xlines-relative}
\end{figure*}

The second row in Figure \ref{fig-xlines-rsmet} demonstrates the changes in line intensities, when irradiated by a young (30 Myr) SSP with the luminosity $L_\ast=3\times 10^9L_\odot$ (the equivalent mass is $M_\ast\simeq 10^{10}\msun$ for a Salpeter IMF). Being strongly suppressed by stellar continuum in $\lambda=0.1\hbox{--}1~\mu$m range, the relative intensities of nebular optical lines drop by two orders of magnitude in the low end of the BH X-ray luminosity. The decrease factor remains nearly constant when the SSP luminosity grows: $L=3\times 10^{10}L_\odot$ (third panel from top, left column), and $=3\times 10^{11}L_\odot$ (lowest panel, left column). This approximate constancy stems from the fact that while the BH with masses $M_\bullet\simlt 10^5\msun$ contributes less towards nebular luminosity in $\lambda\sim 0.1\hbox{--}5\mu$m than the stellar continuum, the nebular emission grows nearly proportionally with stellar luminosity. The intensities of optical lines grow with the BH mass, (nearly  $\propto M_\bullet^{\alpha}$, with $\alpha\approx {1 \over 2}\hbox{--}{1 \over 3}$), for $M_\bullet\leq 3\times 10^6\msun$ for $L_\ast=3\times 10^{10}$ and $=3\times 10^{11}L_\odot$, respectively. 

Mid-IR lines (dark grey shades) in the middle {column} panels show a behaviour similar to the optical (H$\alpha$, H$\beta$ and B$\alpha$) lines, though with a noticeable steepening of growth ${\cal F}_\lambda(M_\bullet) \propto M_\bullet^\alpha$ with $\alpha\sim 1/3,~2/3,~1$ from SSP luminosity $L_\ast=3\times 10^9$ to $L_\ast=3\times 10^{11}\msun$, correspondingly. Far-IR lines of high ionization species (light-dark shades) in the middle columns show a rather weak dependence on BH mass $\alpha\sim 1/6$ to 1 within $M_\bullet\simlt 3\times 10^6\msun$ in the whole range of SSP luminosities, from upper to lower rows of panels. Given this `conservative' trend, the ratios of intensities of hydrogen optical lines, mid-IR lines of the ions [MgVII] and [NeVI] as well as of far-IR lines of low ionization  and neutral species to the lines of [NIII] and [OIII] can serve to measure the ratio of the BH mass (luminosity) to the young SSP luminosity within the {range} shown in Figure \ref{fig-xlines-rsmet} ({ for $M_\bullet\simlt 3\times 10^6\msun$}).

\begin{figure*}
\center
\includegraphics[width=16.0cm]{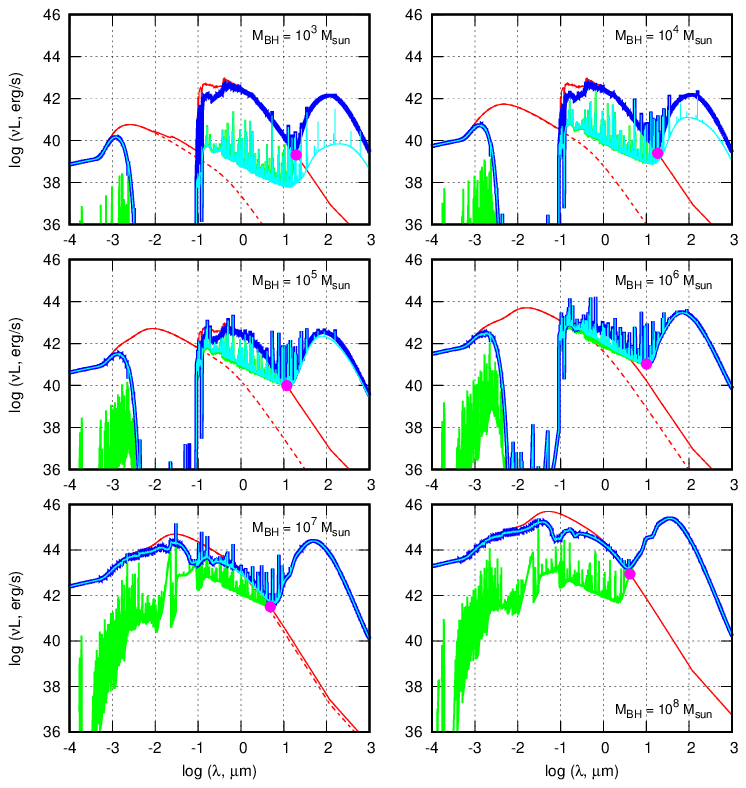}
\caption{
{
Same as in Figure \ref{fig-spec-sspbh} for gas with metallicity ${\rm [Z]}=-1$, and dust with the mass fraction ${\cal D}=0.3\zeta$, the stellar population with  metallicity $-1$, with the luminosity $L_\ast=3.1\times 10^9L_\odot$ and the age of 200~Myr is assumed. Cian lines show the nebular emission under ionization and heating only from the BH. 
}
}
\label{fig-xlines-met1-dst}
\end{figure*}

\begin{figure*}
\center
\includegraphics[width=16.0cm]{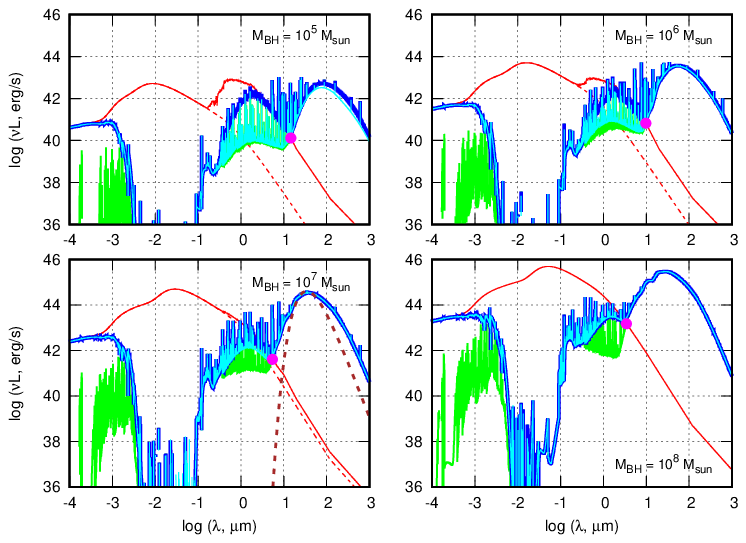}
\caption{Same as in Figure \ref{fig-spec-sspbh} {but only $M_{\bullet} = 10^3$ and $10^4~\msun$} for gas with solar metallicity [Z]=0, and dust with the mass fraction ${\cal D}=0.3\zeta$. Cian lines show the nebular emission under ionization and heating only from the BH. {\it Brown} dashed { line on the lower left panel shows} modified blackbody spectra with spectral index $\beta=2$ and temperature $T=70$ K (see discussion in text). {A stellar population has age 1~Gyr and solar metallicity.} A worthwhile details can be observed in the {upper} left  panel: the presence of gas results in a decrease of the X-ray emission at $\lambda\sim 20$ \AA,\ and an increase of the nebular and dust emission at $\lambda\simgt 10~\mu$m -- such a reversed reaction in the X-ray high energy band and in FIR/submm range encrypt the interrelation between the masses of BH, gas and dust (discussion in Sec.~\ref{disc}).}
\label{fig-xlines-smet-dst}
\end{figure*}

The ratios of intensities between the lines from the three bands -- optical, mid-IR and far-IR/submm to [NIII] 57$\mu$m -- versus the BH mass are presented in Figure \ref{fig-xlines-relative}, for several values of the SSP luminosities from $L=3\times 10^8L_\odot$ to $3\times 10^{11}L_\odot$. The relative intensities of [B$\alpha$/NIII], [NeVI/NIII] and [CIV/NIII] vary in a wide range spanning more than two orders of magnitude, with [B$\alpha$/NIII] and [NeVI/NIII] growing from $\sim 0.03$ to $\sim 3$ when BH mass increases from $\sim 10^3\hbox{--}10^4\msun$ to $\sim 2\times 10^6\msun$, depending on the SSP luminosity. At the same time, [CIV/NIII] drops from $\sim 10^{1.5}\hbox{--}10^{2.7}$ to $<3$ in the same range. Note that the line ratios are not single-valued functions of the $M_\bullet$ to $M_\ast$ ratio. 

As seen from Figure \ref{fig-xlines-rsmet}, for $M_\bullet\simgt 3\times 10^6\msun$ all line intensities decrease steeply because of a growing contribution of the BH  beyond the optical band. Another consequence is that BHs with masses in this range outshine stellar light even with $L_\ast\sim 3\times 10^{11}L_\odot$, so that stellar population becomes difficult to identify. Only the submm continuum nebular emission {becomes the sole carrier of} information about the ionized gas, as shown in the right column of Figure \ref{fig-spec-int}.    

The results discussed above and illustrated in Figures \ref{fig-xlines-rsmet} and \ref{fig-xlines-relative} weakly depend on gas mass, and therefore the intensity ratios [B$\alpha$/NIII], [NeVI/NIII] and [CIV/NIII] can serve as direct measures of masses of the stellar population and the BH feeding the nebular line emissions, unless the BH mass exceeds $M_\bullet\sim 3\times 10^6\msun$. The results however change with gas metallicity: the line intensities of heavy elements grow approximately linearly with metallicity (${\cal F}_\lambda\propto Z$) as shown in Figure \ref{fig-xlines-rsmet-z} in Appendix~\ref{appa}. The intensities of Hydrogen recombination lines slightly decrease with metallicity as $\propto Z^{-1/4}$ because of a decrease in the recombination rates due to a metal enhanced radiative cooling (see left column in Figure \ref{fig-xlines-rsmet-z}). This results in a nearly proportional increase in the intensity ratio [B$\alpha$/NIII] with $Z$ at given $M_\bullet$ and $M_\star$, leaving the ratios [NeVI/NIII] and [CIV/NIII] unchanged. 

\subsection{Signatures of stellar and BH in thermal dust emission} \label{dst_ht}
As seen in Sec. \ref{xsubmm}, the presence of metals in the gas layer clearly manifests itself in a characteristic line spectrum, with line intensities proportional to the metallicity, which can be a useful tool as a diagnostic of the mutual evolution of growing black holes and stellar evolution in a host galaxy, unless dust overwhelms in optical/UV  extinction  and thermal IR emission. 

{However, the presence of dust particles in the interstellar gas complicates the picture.} Assuming a Milky Way dust extinction law with $\sigma_{v}\simeq 5\times 10^{-22}\zeta$ cm$^2$~per H atom \citep{draineb}, the extinction depth of the  gas layer is $\tau_v\simeq 7\zeta M_{g,10}^{1/3}n^{2/3}$, where $\zeta=Z/Z_\odot$ is a metallicity scaling factor, and  $M_{g,10}$ is the gas mass in our fiducial model in the unit of $10^{10}\msun$. In the longer wavelength domain up to $\lambda\simlt 1000~\mu$m, the extinction cross-section decreases as $\sigma(\lambda)\propto \lambda^{-4/3}$, for our estimates we assumed the approximation $\sigma\approx 5 \times 10^{-22} \zeta (\lambda/0.5 \, \mu{\rm m})^{-4/3}$ cm$^2$ \citep[see Fig. 5 in ][]{compiegne11}. As a result, the gas layer becomes transparent, $\tau(\lambda)\sim 1$, in the near IR and beyond $\lambda\sim 2\zeta^{3/4}M_{g,10}^{1/4}n^{1/2}~\mu$m.   

Dust is heated by UV and X-ray radiation from stars and the BH. A considerable fraction of energy absorbed by ambient gas, as seen in {Figures~\ref{fig-xlines-met1-dst} and \ref{fig-xlines-smet-dst}}, transforms into ionization and heating of the gas and dust. This energy is then partly irradiated in nebular and dust emission. Contribution of the BH dominates gas and dust heating when its mass $M_\bullet\simgt 10^5\msun$ -- it is shown by cyan lines in {Figures~\ref{fig-xlines-met1-dst} and  \ref{fig-xlines-smet-dst}} (which are similar to Figure \ref{fig-spec-sspbh} but for higher metallicity gas, with $[Z]=-1$ and $0$) . It is also seen that not only the absolute value of dust emission but even its fraction -- the ratio of dust luminosity to the sum of luminosity of stars and BH absorbed in the gas layer--- increases with the BH mass, as illustrated in the {\it lower panel} of Figure \ref{fig-xe-te}. 

{We also find that for  $\zeta\geq 0.1$, the optical depth $\tau_{x,uv}$ of the gas volume with radius $r_h\sim (3M_g/4\pi\rho)^{1/3}\simeq 1.5\times 10^{22}M_{g,10}^{1/3}n^{-1/3}$ cm  in X-ray and UV bands exceeds one. Therefore for a rough estimate of dust temperature, one can assume that the most  of the heating radiation is absorbed by dust}. We also assume the Planck-averaged  absorption efficiency $Q(a,T_d)\simeq \langle Q(a,\nu)\rangle\simeq 0.1 aT_d^2$, \citep{dwek92,draineb}, where $a$ is the grain size in cm, and also that the dust spectral index $\beta=2$: $Q_\nu(T_d)\propto\nu^\beta$. With these assumptions the dust temperature is given by
\be \label{td}
T_d\sim 17 \,{\rm K} (\eta_h\omega)^{1/6}\left({\langle Q_{v}\rangle L_{\ast,9}+\langle Q_{x}\rangle L_{\bullet,9}\over a_{0.1}}\right)^{1/6}\left({n\over M_{g,10}}\right)^{1/9}
\ee
where $a_{0.1}$ is the dust radius $a$ in 0.1$\mu$m, $L_{\ast,9}$ and $L_{\bullet,9}$ are the stellar and black hole luminosities in $10^{9}L_\odot$, $M_{g,10}$ the gas mass in $10^{10}\msun$, $\eta_h=0.1\hbox{--}0.3$ (see {\it Lower panel} on Figure \ref{fig-xe-te}) is the fraction of energy from the sources -- stars and BH, that is used in dust heating, $\omega$ is the mean solid angle of X-ray and UV photons impinging on the dust grain. In estimates, the absorption efficiencies of dust particles in visual ($v$), EU and X-ray ($x$) bands are assumed to be $\langle Q_{v,x}\rangle\sim 1$ \citep[see, Fig. 24.1 in ][]{draineb}. This implies that the characteristic wavelength at the peak dust emission is 
\be \label{lamtd}
\lambda_{T_d}\sim 170\left({a_{0.1}\over \omega\eta_h\Sigma L_{9}}\right)^{1/6}\left({M_{g,10}\over n}\right)^{1/9}\mu{\rm m}. 
\ee
with $\Sigma L_{9}=\langle Q_{abs}(v)\rangle L_{\ast,9}+\langle Q_{abs}(v)\rangle L_{\ast,9}$. {The estimate is roughly consistent with the peak position in Figure \ref{fig-xlines-met1-dst}.}

The $T_d$ in Eq. (\ref{td}) is obviously a lower estimate of the dust temperature, {whereas in reality, a large fraction of dust is at a higher temperature,} predominantly contributing into the peak at $\lambda\sim 70\mu$m. This hot dust is associated with the internal layers with $\tau_v\sim 7$ for UV and visual photons, and $\tau_x\sim 3$ for X-ray photons with energy $\epsilon\sim 1$ keV, where most of heating and ionizing radiation from the stars and the BH is absorbed.  This situation is also reflected in a very weak dependence of $\eta_h$ on the gas mass on {\it lower panel} of Figure \ref{fig-xe-te}. Correspondingly, the peak wavelengths in (\ref{lamtd}) should be treated as a {upper} limit of the peak wavelength. {This explains the fact  that} the dust emission in Figure \ref{fig-xlines-smet-dst} cannot be fit by an isothermal (single temperature) dust. In the left column lowermost panel of Figure \ref{fig-xlines-smet-dst}, we show {a modified Planck curve with the emissivity spectral index $\beta=2$ and temperature 70 K for $M_\bullet=10^7\msun$} {by thick dashed brown line}. A deficit of emission (of factor $\sim 3$ at $\lambda\sim 100~\mu$m to nearly two orders at $\sim 1000~\mu$m in the lower panel) is clearly seen between the modified Planck spectrum and the modeled one, which reflects the negative radial gradient of the dust temperature.         

With this stipulation, an upper limit of the dust luminosity can be estimated if one assumes that the fraction $\eta_h$ of the total (stellar and BH) luminosity heats the {\it entire} dust mass in the galaxy: $M_d={\cal D}Z_\odot M_g$ with ${\cal D}Z_\odot$ being the mass fraction of dust. Assuming the average  input rate of heating radiation impinging a dust grain to be $\sigma\omega\eta_h\Sigma L/4\pi R^2$ and the total number of dust particles in the galaxy $N_d=M_d/m_d$, with $m_d=4\pi\rho_d/3a^3$, and $M_g=4\pi\rho R^3/3$, one arrives then at,
\be\label{dlum}
L_d\sim \eta_h\omega L_9{{\cal D}Z_\odot\rho\over 4\pi a\rho_d}R,
\ee
with $\rho_d=3$ g cm$^{-3}$ being the dust grain density. For ${\cal D}=0.3\zeta$, {$\zeta=0.1$, and $Z_\odot=0.02$ one arrives at 
\be \label{heat}
L_d\sim 2\times 10^{41}\eta_h\omega(L_{\ast,9}+L_{\bullet,9})M_{g,10}^{1/3}n^{2/3}~{\rm erg~s^{-1}}.
\ee
The} contribution from the BH in dust heating obviously prevails  when its bolometric luminosity becomes larger than the stellar bolometric luminosity $L_\bullet=L_\ast$. For the model shown in Figure \ref{fig-xlines-met1-dst}, it occurs {when $M_\bullet\simgt 10^6\msun$, assuming} the critical (Eddington) accretion rate. Beyond this limit information about the stellar population cannot be retrieved from the net spectrum.

At lower masses {of BHs $M_\bullet<10^6\msun$, contributions} from BH and stars can be distinguished by making use of the hard X-ray emission of the BH. One can estimate from Figures \ref{fig-spec-sspbh} and \ref{fig-xlines-met1-dst} that the luminosity in the hard X-ray (energies $E_X\sim 3$ keV) varies approximately as $L_x\propto M_\bullet^{4/5}$). The luminosity of dust emission grows as $L_{d,p}\propto (M_\ast+M_\bullet)$. Combining the observed fluxes of the hard X-ray and the dust peak emission one can infer separately $M_\ast$ and $M_\bullet$ unless $M_\bullet\simgt 10^5\msun$. 

More specifically, the two equations connecting the X-ray and thermal dust observables are: $L_x\simeq 5\times 10^{36}M_\bullet^{4/5}$ erg~s$^{-1}$ {and $L_{d}\simgt 2\times 10^{41}(L_{\ast,9}+3\times 10^{-5}M_\bullet)$ erg~s$^{-1}$ as} follows from Eq. (\ref{heat}). Here the second term in the parenthesis is  the Eddington luminosity $L_{\rm Edd}=1.24\times 10^{38}M_\bullet$ normalized by $10^9L_\odot$, and $M_\bullet$ is in $\msun$ unit. As a result, 
\be 
M_\bullet\simeq 10^5\left({L_x\over 5\times 10^{40} \, {\rm erg} \, {\rm s}^{-1}}\right)^{5/4}, 
\ee
{and 
\be \label{stell}
L_{\ast,9}\simlt 28\eta_h^{-1}M_{g,10}^{-1/3}n^{-2/3}L_{d,9}-7\left({L_x\over 10^{41}\, {\rm erg} \, {\rm s}^{-1}}\right)^{5/4}\,.
\ee
Here the} luminosities $L_{d}$ and $L_{\ast}$ are normalized by their fiducial values in $10^9L_\odot$. 
Note that Eq. (\ref{stell}) is valid {for $L_x\leq 2\times 10^{42}M_{g,10}^{-4/15}n^{-8/15}L_{d,9}^{4/5}$ erg~s$^{-1}$, i.e. it is within the limits of the BH contribution in dust heating not exceeding the stellar one, or, equivalently, $M_\bullet\leq 10^6\msun$)}. 

\subsubsection{Spectral lines in a dusty environment}

{Galaxies with higher stellar and gas metallicities [Z/H]=0 are supposed to be further evolved, and as a consequence, to be  more massive, harbouring more massive black holes. In Figure \ref{fig-xlines-smet-dst} we show the spectra for galaxies with an age of $1$ Gyr, stellar luminosity $L_\ast=3.1\times 10^9~L_\odot$, $M_g=10^{10}~\msun$ and metallicity [Z/H]=0. Panels from left uppermost to right lowermost show cumulative spectra with black holes of masses $M_\bullet=10^5,~10^6,~10^7,~10^8~\msun$, respectively.} 

{In a high-metallicity gas spectral lines are obviously more pronounced as compared to those in low-metallicity gas (see Figure~\ref{fig-xlines-rsmet-z}). However, dust emission floods into the IR/submm range $\lambda\geq 3\hbox{--}10~\mu$m, and the majority of lines become immersed into the dust emission (Figures~\ref{fig-xlines-met1-dst} and \ref{fig-xlines-smet-dst}). Therefore, in presence of dust, only a few of relatively strong spectral lines in optical and near-IR range are available for $M_\bullet \simgt 10^4~\msun$. For gas metallicity  $\rm [Z/H] = -1$ spectral lines are marginally distinguishable only for black hole masses $10^5\leq M_\bullet\leq 10^6~\msun$, and can hardly serve to measure the interrelations between the masses of BH, gas and stellar population. 
Spectral lines in a gas with higher metallicity are stronger and a few of them remain distinguishable up to $M_\bullet \sim 10^8~\msun$ (Figure~\ref{fig-xlines-smet-dst}). These are: [FeII]~5.3$\mu$m from regions of low ionization, and [NeVI]~7.6$\mu$m, [MgVII]~5.5$\mu$m, 9$\mu$m in from high ionization regions. The ratios of their relative luminosities ${\cal L}_{\rm NeVI}=(L_{\rm NeVI}-L_c)/L_c$,  ${\cal L}_{\rm MgVII,5.5}=(L_{\rm MgVII,5.5}-L_c)/L_c$ and  ${\cal L}_{\rm MgVII,9}=(L_{\rm MgVII,9}-L_c)/L_c$ vary with $M_\bullet$ and $M_g$ and might be utilised to measure them. However, their interrelation becomes degenerate in the higher limit of gas mass $M_g\simgt 3\times 10^{9}~\msun$, and in addition does not form a regular sequence along the stellar luminosity.   
}

\begin{figure}
\center
\includegraphics[width=8cm]{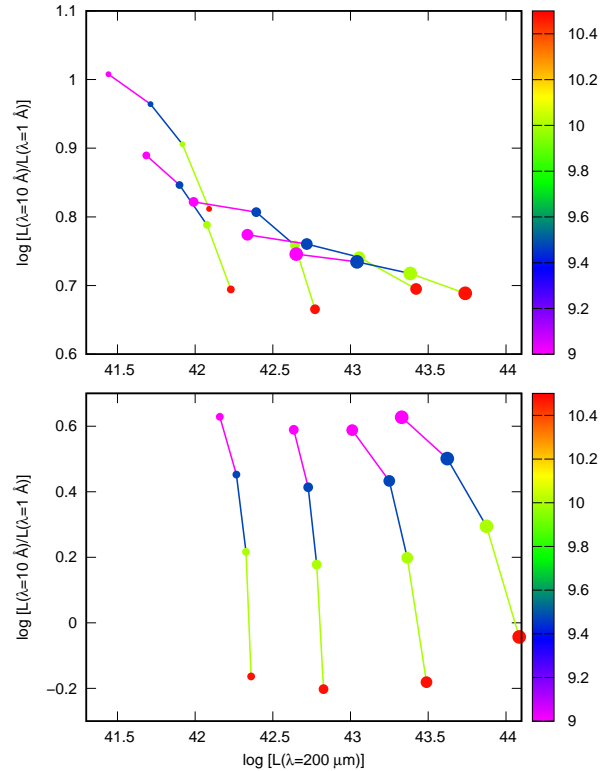}
\caption{{Dependence of the ratio of X-ray emission at 10\AA\, to 1\AA\, on the sub-mm continuum emission of dust and ionized gas at 200 $\mu$m for different  masses of the BH: from bottom to top $10^4,~10^5,10^6,~10^7,~10^8~\msun$, as indicated by growing sizes of circles; colours show the gas mass in solar units as coded in the colour bar. {Upper panel:} [Z/H]=$-1$, {\it lower panel:} [Z/H]=0. It is  seen that in case of lower metallicity the ratio varies with gas mass more weakly than for [Z/H]=0. The results are shown for fiducial stellar luminosity {$L_\ast=3.1\times 10^9~L_\odot$}; the dependences $L(10$\AA$)/L(1$\AA$)$\, vs $L(200~\mu$m$)$ remain essentially invariant for stellar luminosities} $L_\ast \simlt L_{BH}$. In case of higher $L_\ast$, the correlations shift to the right, such that at $L_\ast=3\times 10^{10}~L_\odot$ the shift for BH mass $M_{BH} \sim 10^5\msun$ ($L \sim 3\times 10^9\lsun$) is around an order of magnitude towards higher emission at $200~\mu$m. 
}
\label{fig-lambda-1}
\end{figure}

\section{Discussion} \label{disc}

{The main emphasis of the present paper is on finding a method of determining the masses of three important constituents of high redshift galaxies --- central supermassive black hole, stars and gas --- with the aid of spectral information in X-ray, optical, mid and far infrared wavelengths.}

\begin{itemize} 
 \item Observations of very distant ($z>6$) galaxies hosting growing supermassive black holes provide only their spectra without spatial (angular) structure, and do not allow resolving stellar population and the black hole separately. Therefore, in such conditions, estimates of their masses can only be made from the spectra. Hovewer, even though spectral features of relatively young (with the age less than 200 Myr) stellar population are to a certain extent distinguishable from those of black holes, the presence of gas and dust of unknown masses may complicate the situation, thus making a separate determination of masses of stellar population and the hosted SMBH challenging.    
 
 \item A possible way to overcome this difficulty may involve measurements of emissions that can separately characterize the contributions from the stellar population, the black hole, the gas nebular emission and the dust emission: {\it i)} high energy X-ray emission $E\sim 1\hbox{--}10$ keV can come mostly from the central BH, {\it ii)} thermal dust emission is supported by heating from the BH and the stars, {\it iii)}  far IR nebular continuum emission in $\lambda\sim 200\mu$m can give information about the gas mass. With such a multi-wavelength approach, the three observables would provide us with necessary information to infer the three unknowns. Presence of dust complicates such a robust approach and urges additional consideration.  
 
 \item The role the gas component plays in shaping the cumulative spectrum of a host galaxy bearing a growing black hole is dual: one is that gas absorbs X-ray, UV and optical emission from the BH and stars, the other is that nebular emission contains spectral lines, dust continuum and free-free continuum in far IR and submm. One can therefore expect that increasing of the gas mass suppresses X-ray and UV/optical light from the sources, and simultaneously enhances nebular emissions. Such a trend is clearly seen in the left {upper} panel of Figure \ref{fig-xlines-smet-dst}: the solid red line depicts the cumulative spectrum of stars and BH without absorption and emission from gas. As soon as these effects are turned on, they immediately decrease the amount of X-ray and UV photons -- the higher the gas mass, the lower is  the ionizing photon energy -- and of UV/optical/NIR photons in the range from $\lambda\sim 1000$ \AA\, to $\lambda\sim 2\mu$m. The diagram on Figure \ref{fig-lambda-1} show such an inverse dependence of an increased of submm free-free $200~\mu$m flux and a decrease of the ratio of X-ray fluxes at $10$\AA\, to $1$\AA\, along with an increasing gas mass. 
 
 \item This trend can be observed particularly at the high end of BH masses in Figure \ref{fig-lambda-1}. It is seen that with an increase in the gas mass,  the 200~$\mu$m flux enhances, with a simultaneous decline of the $10$\AA\, to $1$\AA\, fluxes. Such a concerted anti-correlation of 200~$\mu$m and $10$\AA\, to $1$\AA\,  fluxes with a growing gas mass can also be seen at smaller BH masses, though  becoming flatter with smaller the BH masses. This is because of the  fact that the ionizing radiation flux decreases as $\exp(-\tau_x)$,  where $\tau_x\propto N_{\rm H}\propto M_g^{1/3}$ is the optical depth in X-ray/UV (for the assumed parameters $\tau_x\simgt 1$), while the  200~$\mu$m flux grows as a shallow power-law scaling $\propto M_g^{1/3}$. 
 
 \item  In our consideration we explicitly used the black hole spectral energy distribution based on the model described by \citet{kubota18}, with the X-ray  luminosity ($\epsilon=0.03\hbox{--}10$ keV) scaling as $L_x\propto M_\bullet^{4/5}$. Observationally the interrelation between the X-ray luminosity and the BH mass seems challenging to be robustly inferred, mostly because of contamination from heavy dust obscuration, or lack of direct measures of BH mass when X-ray emission is recognizable \citep[see discussion in ][]{circosta19,connor19,vito19,lambrides20,damato20}. However, our results are qualitatively correct for any positive power-law dependence $L_x\propto M_\bullet^\alpha$, $\alpha>0$, which is obviously the case. Our arguments and resulting interrelations remain largely invariant in this case.
\end{itemize}

\section{Conclusions}

\noindent
We have studied observational manifestations of the interrelation between the masses of stellar population, central massive black hole and the interstellar gas in the  host galaxies through their multiwavelength spectral energy distributions from X-ray to submm wavelengths. We modelled cumulative spectra as a sum of emission from growing BHs, stellar populations and nebular emission from interstellar and circumgalactic mass with making use of CLOUDY (version {17}). We argue that the three observables: the luminosities (fluxes) in X-ray, dust infrared and submm thermal emission, along with their ratios allow us to infer the masses (luminosities) of the stellar population, the central BH and the gaseous component.     

Our results are summarized as follows:  

\begin{itemize}
 \item In dust-free models, emission of growing black holes dominates when $M_\bullet/M_\ast\geq 0.02$. In this case nebular free-free continuum in the far infrared outshines the emission from central source, with a change of spectral shape from a quasi-blackbody spectrum $\propto\nu^2$ to a flat $\propto\nu^{-0.12}$ free-free continuum. The frequency $\nu_k$ and the corresponding luminosity $L_k$ at which the spectral index changes, depend on the BH and stellar population masses, and also on the gas mass. Thus, the observed $\nu_k$ and $L_k$ provide two interrelations connecting the three masses: $M_\bullet$, $M_\ast$ and $M_g$. 
 
 \item Measuring the mass of stellar population is possible only for a relatively low mass of BH $M_\bullet\simlt 3\times 10^6\msun$, otherwise stellar emission is totally out shined by the BH emission. 
 
 \item Line ratios are not single-valued functions of the mass ratio $M_\bullet/M_\ast$, and thus cannot serve to measure this ratio.      
 
 \item The thermal dust emission in its peak around 70 $\mu$m and the hard X-ray emission at $\lambda\sim 2\hbox{--}4$\AA\, allow us to measure masses of the stellar population and the black hole.   
 
 \item Interrelations between the fluxes (luminosities) of X-ray photons with $E\sim 3$ keV and submillimeter emission at $\lambda\sim 200\mu$m allow to infer the gas mass in the galaxies, where the BH mass is $M_\bullet>10^5\msun$ and the stellar population luminosity $L_\star/L_\odot<3\times 10^4(M_\bullet/\msun)$. 
 \item The dependences of the ratio $L(10$\AA$)/L(1$\AA$)$\, vs $L(200~\mu$m$)$ traces the interrelation between the dust and gas masses, though they remain essentially invariant for stellar luminosities $L_\ast \simlt L_{BH}$. In case of higher $L_\ast$, the correlations shift to higher submm luminosity emission at $200~\mu$m.

\end{itemize}

\section{Acknowledgements}

\noindent

This work is supported by the joint RFBR-DST project (RFBR 17-52-45063). EV is grateful to the Ministry for Education and Science of the Russian Federation (grant 3.858.2017/4.6). The work by YS is done under partial support from the project 01-2018 ``New Scientific Groups LPI''.

    \bibliographystyle{mn2e}

\begin{thebibliography}{} 


\bibitem[\protect\citeauthoryear{Abramowicz \etal}{1988}]{abram88} Abramowicz, M. A., Czerny, B., Lasota, J. P., Szuszkiewicz, E., 1988, ApJ, 332, 646   

\bibitem[Agarwal \etal(2013)]{agarwal13} Agarwal, B., Davis, A. J., Khochfar, S., \etal 2013, MNRAS, 432, 3438 

\bibitem[Bell \& de Jong(2001)]{bell01} Bell, E. F., de Jong, R. S., 2001, ApJ, 550, 212

\bibitem[Bentz \etal(2013)]{bentz13} Bentz, M. C., Donney, K. D., Grier, C. J., \etal 2013, ApJ, 767, 149 

\bibitem[Bentz \& Manne-Nicholas(2018)]{bentz18} Bentz, M. C., Manne-Nochilas, E., 2018, ApJ, 864, 146 

\bibitem[Bruzual \& Charlot(2003)]{bruz03} Bruzual, Charlot, 2003 

\bibitem[Circosta \etal(2019)]{circosta19} Circosta, C., Vignali, C., Gilli, R., \etal 2019, A\&\,A, 623, A172

\bibitem[Compi\`egne \etal(2011)]{compiegne11} Compi\`egne, M., Verstraete, L., Jones, A., \etal, 2011, A\&\,A, 525, A103  

\bibitem[Connor \etal(2020)]{connor19} Connor, T., Ba\~nados, E., Stern, D., \etal 2020, ApJL, 884, L31 

\bibitem[D\'Amato \etal(2020)]{damato20} D\'Amato, Q., Gilli, R., Vignali, C., \etal 2020, arXiv:2003.08631

\bibitem[Decarli \etal(2018)]{decarli18} Decarli, R., Walter, F., Venemans, B. P., \etal 2018, ApJ, 857, 97 

\bibitem[Draine(2011)]{draineb} Draine, B. T., Interstellar and Intergalactic Medium 

\bibitem[Dwek \& Arendt(1992)]{dwek92} Dwek, E., Arendt, R. G., 1992, ARA\&\,A, 30, 11

\bibitem[Faber \& Gallagher(1979)]{faber79} {Faber, S. M., Gallagher, J. S., 1979, ARA\&\,A, 17, 135 }

\bibitem[Ferland \etal(2017)]{cloudy17} Ferland, G.J., Chatzikos, M., Guzm\'an, F. \etal 2017, Rev. Mex. de Astron. Astrofis., 53, 385 

\bibitem[Girardi \etal(2000)]{padova2000} Girardi L., Bressan A., Bertelli G., Chiosi C., 2000, A\&AS, 141, 371

\bibitem[Graham \& Scott(2013)]{graham13} Graham, A. W., Scott, N., 2013, ApJ, 764, 151 

\bibitem[H\"aring \& Rix(2004)]{haring04} H\"aring, N., Rix, H. W., 2004, ApJL, 604, L89 

\bibitem[Heckman \& Best(2014)]{heckman14} Heckman, T. H., Best, P. N., 2014, ARA\&\,A, 52, 589  

\bibitem[Into \& Portinari(2013)]{into13} Into, T., Portinari, L., 2013, MNRAS, 430, 2715 

\bibitem[Kauffmann \etal(2003)]{kauf03} Kauffmann, G., \etal 2003, MNRAS, 341, 33 

\bibitem[Kormendy \& Ho(2013)]{kormen13} Kormendi, J., Ho, L. C., 2013, ARA\&\,A, 51, 511 

\bibitem[Kubota \& Done(2018)]{kubota18} Kubota, A., \& Done, C. 2018, MNRAS, 480, 1247 

\bibitem[Latif \& Ferrara(2016)]{latif16} Latif, M. A., Ferrara, A., 2016, PASA, 33, 51 

\bibitem[Lambrides \etal(2020)]{lambrides20} Lambrides, E., Chiaberge, M., Heckman, T., \etal 2020 arXiv:2002.00955 

\bibitem[Lupi \etal(2016)]{lupi16} Lupi, A., Haardt, R. Dotti, M., 2016, MNRAS, 456, 2993 

\bibitem[Madau \etal(2014)]{madau14} Madau, P., Haardt, F., Dotti, M., 2014, ApJ, 784, L38 

\bibitem[Marconi \& Hunt(2003)]{marconi03} Marconi, A., Hunt, L. K., 2003, ApJL, 589, L21  

\bibitem[Natarajian(2014)]{natarajian14} Natarajan, P., 2014, GReGr, 46, 1702  

\bibitem[Natarajan \etal(2017)]{nataraj17} Natarajan, P., Pacucci, F., Ferrara, A., \etal 2017, ApJ, 838, 117 

\bibitem[Novikov \& Thorne(1973)]{novik73} Novikov, I. D., Thorne, K., 1973, in: {\it Black Holes}, eds. C. de Witt \& B. S. de Witt, Gordon \& Breach, p. 345 

\bibitem[Pacucci \etal(2016)]{pacucci16} Pacucci, F., Ferrara, A., Grazian, A., \etal 2016, MNRAS, 459, 1432 

\bibitem[Pacucci \etal(2015)]{pacucci15} Pacucci, F., Volonteri, M., Ferrara, A., 2015, MNRAS, 452, 1922 

\bibitem[Pezzulli \etal(2016)]{pezzulli16} Pezzulli, E., Valiante, R., Schneider, R., 2016, MNRAS, 458, 3047 

\bibitem[Reines \& Volonteri(2015)]{reines15} {Reines, A. E., Volonteri, M., 2015, ApJ, 813, 72 }

\bibitem[Salim \etal(2005)]{salim05} Salim, S., Charlot, S., Reach, A. M., \etal 2005, ApJ, 619, L39 

\bibitem[Sadowski \etal(2009)]{sadow09} Sadowski, A., 2009, ApJSS, 183, 171 

\bibitem[Sani \etal(2011)]{sani11} Sani, E., Marconi, A., Hunt, L. K., Risaliti, G., 2011, MNRAS, 413, 1479  

\bibitem[Tanaka \& Haiman(2009)]{tanaka09} Tanaka, T., Haiman, Z., 2009, ApJ, 696, 1798 

\bibitem[Valiante \etal(2016)]{valiante16} Valiante, R., Schneider, R., Volonteri, M., Omukai, K., 2016, MNRAS, 462, 3146 

\bibitem[Valiante \etal(2018)]{valiante18} Valiante, R., Schneider, R., Zappacosta, L., \etal 2018, MNRAS, 476, 407 

\bibitem[Vito \etal(2019)]{vito19} Vito, F., Brandt, W. N., Bauer, F. E., \etal 2019, A\&\,A, 2628, L6 

\bibitem[Volonteri(2012)]{volon12} Volonteri, M., 2012, Sci.? 337, 544 

\bibitem[Volonteri \etal(2015)]{volonteri15} Volonteri, M., Silk, J., Dubus, G., 2015, ApJ, 804, 148 

\bibitem[Volonteri \& Reines(2016)]{volont16} Volonteri, M., Reines, A. E., 2016, ApJ, 820, L6 

\bibitem[Volonteri \etal(2016)]{volonterid16} {Volonteri, M., Dubois, Y., Pichon, C, Devriendt, J., 2016, MNRAS, 460, 2979}

\bibitem[Volonteri \etal(2017)]{volont17} Volonteri, M., Reines, A. E., Atek, H., \etal, 2017, ApJ, 849, 155 

\bibitem[Wilms \etal(2000)]{wilms00} {Wilms, J., Allen, A. McCray, R., 2000, ApJ, 542, 914} 

\bibitem[Woo \etal(2013)]{woo13} Woo, J.-H., Schulze, A., Park, D. \etal 2013, ApJ, 772, 49  

\end{thebibliography}

\appendix

\section{{Intensities versus metallisity}} 
\label{appa}

\begin{figure*}
\center
\includegraphics[width=16cm]{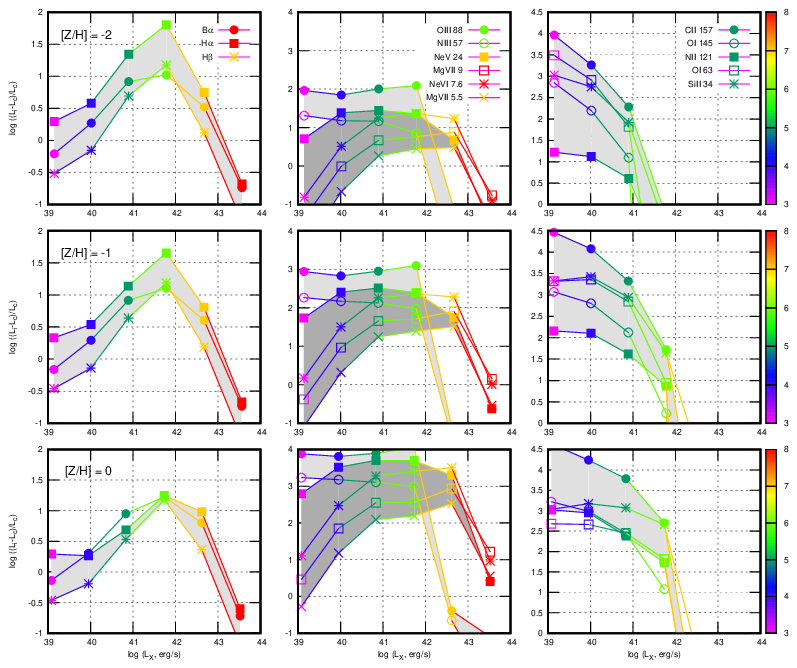}
\caption{The line intensities as in Figure \ref{fig-xlines-rsmet} for different gas metallicity: $\rm [Z/H]=-2,\ -1,$ and 0, from top to bottom. }
\label{fig-xlines-rsmet-z}
\end{figure*}

Increase of gas metallicity, under other conditions being equal, enhances radiative cooling and shifts thermal equlibrium to a slightly lower temperature. As a results recombination rates increase and result in a decrease of fractional ionization. This results in turn in a weakening of hydrogen recombination lines $\propto Z^{-1/4}$. It is demonstrated in left panels on Figure \ref{fig-xlines-rsmet-z}. On the contrary, middle and right panels show that optical, infrared and submm lines of the ions of heavy metals rise nearly proportional with $Z$. {Note that here we use the SED for a young (30~Myr) stellar population for all gas metallicity values.} 

\section{{Intensities versus gas mass}} 
\label{appb}

\begin{figure*}
\center
\includegraphics[width=16cm]{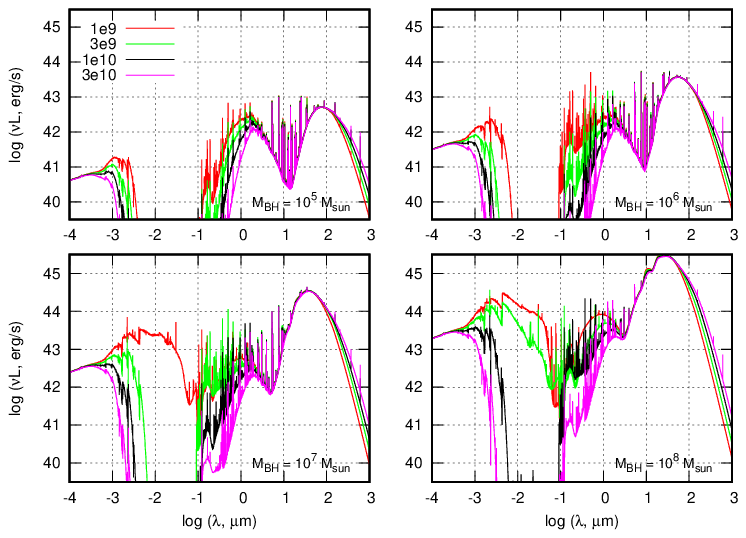}
\caption{
Nebular emission from gas exposed by the stellar population with bolometric luminosity $L_\ast=3\times 10^{9}L_\odot$ and the black hole with masses $M_\bullet = 10^5, 10^6, 10^7$ and $10^8~\msun$ from left top to right bottom, respectively. Different curves depict spectra of gas mass as legended; gas metallicity [Z]=0, dust mass fraction assumed as in the Milky Way ${\cal D}=0.3\zeta$; decrease of X-ray luminosity at $\lambda\geq 10$\AA, and increase of nebular FIR emission $\lambda\simgt 200 \mu$m with an increase of gas mass (from red to magenta lines are clearly seen. {The stellar population has age 1~Gyr and solar metallicity.}
}
\label{fig-spec-sspbh-dust-m}
\end{figure*}

{When the stellar population dominates -- for the Eddington luminosity it is when $M_\bullet/\msun\leq  3\times 10^{-5}(L_\star/L_\odot)$, it overwhelms the nebular emission and the metals spectral lines are uneloquent, as can be seen in Figure \ref{fig-xlines-smet-dst}, and hardly can be used to infer the gas mass.} Situation changes for  $M_\bullet/\msun> 3\times 10^{-5}(L_\star/L_\odot)$ with a well shaped nebular spectrum showing explicit dependence on gas mass. The most interesting feature is an approximately inversely proportional decrease of X-ray emission due to strong absorption seen at $\lambda\sim 10$\AA\,, and an approximately proportional increase of emission in submm waveband at $\sim 200\mu$m with growing gas mass, Figure \ref{fig-spec-sspbh-dust-m}. 

\end{document}